\documentclass[conference]{IEEEtran}
\IEEEoverridecommandlockouts

\usepackage{cite}
\usepackage{amsthm, amsmath,amssymb,amsfonts}
\usepackage{algorithmic}
\usepackage{graphicx}
\usepackage{textcomp}
\usepackage{xcolor}

\usepackage{enumitem}
\usepackage{hyperref}
\usepackage{array}
\usepackage{eqparbox}
\usepackage{caption}

\makeatletter
\g@addto@macro{\UrlBreaks}{\UrlOrds}
\makeatother

\def\BibTeX{{\rm B\kern-.05em{\sc i\kern-.025em b}\kern-.08em
    T\kern-.1667em\lower.7ex\hbox{E}\kern-.125emX}}

\newtheorem{definition}{Definition}
\begin{document}

\title{Information-theoretic Estimation of the Risk of Privacy Leaks\\
}

\author{\IEEEauthorblockN{Kenneth Odoh}
\IEEEauthorblockA{\textit{\url{https://kenluck2001.github.io}}\\
Vancouver, Canada \\
kenneth.odoh@gmail.com}
}

\maketitle

\begin{abstract}

Recent work~\cite{Liu2016} has shown that dependencies between items in a dataset can lead to privacy leaks. We extend this concept to privacy-preserving transformations, considering a broader set of dependencies captured by correlation metrics. Specifically, we measure the correlation between the original data and their noisy responses from a randomizer as an indicator of potential privacy breaches. This paper aims to leverage information-theoretic measures, such as the Maximal Information Coefficient (MIC), to estimate privacy leaks and derive novel, computationally efficient privacy leak estimators. We extend the $\rho_1$-to-$\rho_2$ formulation~\cite{Evfimievski2003}  to incorporate entropy, mutual information, and the degree of anonymity for a more comprehensive measure of privacy risk. Our proposed hybrid metric can identify correlation dependencies between attributes in the dataset, serving as a proxy for privacy leak vulnerabilities. This metric provides a computationally efficient worst-case measure of privacy loss, utilizing the inherent characteristics of the data to prevent privacy breaches.

\end{abstract}

\begin{IEEEkeywords}
Cryptography, Privacy, Security, Data mining
\end{IEEEkeywords}

\section{Introduction}

There is a growing need for privacy-preserving data mining because of the proliferation of privacy legislation and a heightened sense of privacy awareness among customers. Organizations can suffer reputational damage and potentially face class-action suits following privacy breaches. Furthermore, there are numerous examples of firms trading on private customer data. For example, tax data is sensitive information and reported privacy breaches incur heavy penalties, as a recent case suggests in the news~\footnote{\url{https://www.cbsnews.com/news/ex-irs-contractor-charles-littlejohn-trumps-tax-records-sentenced/}}. Furthermore, there is a recent case where tax organizations colluded with social media giants~\footnote{\url{https://www.theverge.com/2022/11/22/23471842/facebook-hr-block-taxact-taxslayer-info-sharing}}. Privacy breaches are on the rise even though they are frequently underreported in the media~\footnote{\url{https://news.umich.edu/data-breaches-most-victims-unaware-when-shown-evidence-of-multiple-compromised-accounts/}} due to the associated costs.

The pervasive issue of privacy abuse has motivated our research to mitigate privacy leaks in data by developing a metric to assess privacy vulnerability. We aim to answer the question: "What is a measurable metric that can quantify the privacy vulnerability risk of data after privacy-preserving transformations?" Previous work~\cite{Reshef2011, Lazarsfeld2022} has shown promise in using mutual-information-based metrics like the Maximal Information Coefficient (MIC) to estimate correlations. Inspired by this approach, we consider correlation a proxy for privacy leak vulnerability. However, these methods can be computationally expensive due to partitioning operations. Our work focuses on creating computationally efficient privacy risk metrics, leveraging correlation as a proxy for potential privacy leaks with real-world applicability. We aim for practical applicability by empirically estimating privacy risk after applying privacy-enhancing transformations. Alternatively, we can use our framework to evaluate the susceptibility to privacy leaks of various privacy-preserving systems in a white-box setting.

The paper is structured as follows: a recap of contributions in Section~\ref{contributions}, a background showing previous work in relation to the current thesis in Section~\ref{related-work}, an overview of our formulation in Section~\ref{extensions}, a discussion in Section~\ref{discussions}~\footnote{\textbf{Source code}: \url{https://github.com/kennex2004/miscellaneous/tree/main/leakdetector}}. Finally, we present limitations, future work, and conclusions in Section~\ref{conclude}. Hence, these terminologies are used interchangeably for the remainder of this work: randomizer, privacy-enhancing mechanism, and transformation depicts a privacy-enhancing mechanism as a scheme that transforms data into a privacy-preserving form that does not leak identifiable information about individual records. 

\subsection{Contributions}
\label{contributions}

Previous work~\cite{Liu2016} has demonstrated that correlations between data samples can compromise privacy in privacy-preserving applications. Building on this insight, we investigate how data correlation can serve as a metric for measuring susceptibility to privacy leaks. Several studies~\cite{Biswas2023} have shown that correlation can be a significant factor in privacy breaches, leading to the development of privacy-aware techniques like Pufferfish~\cite{Kifer2014} that aim to provide privacy even in the presence of correlated data and publicly released data.

This work introduced a novel complexity metric by adapting the degree of anonymity and then extending the $\rho_1$-to-$\rho_2$ formulation~\cite{Evfimievski2003} to enable the use of entropy, mutual information, and degree of anonymity for measuring privacy risk. Our work adopts the $\rho_1$-to-$\rho_2$ framework and utilizes well-studied measures such as degree of anonymity. Our construction utilizes permutation entropy~\cite{Kozak2020} based on the symbolic representation of the patterns in the data, thereby enabling support for various data types (numeric, ordinal, and categorical). Our hybrid correlation metric design considers computational cost, predictive power, interpretability, and ease of use. We utilize information-theoretic measures to provide a domain-independent method to estimate correlations to indicate potential privacy violations.

\section{Related work}
\label{related-work}

Recent studies by Sankhya~\cite{Rainio2022} have provided the characteristics of several correlation metrics. They grouped the dynamics of the underlying dependencies captured by the correlation coefficient, which include linear, non-linear, functional, non-functional, and complex relationships. 

Correlation coefficients not based on information theory include Pearson correlation, Spearman correlation, distance correlation~\cite{Rainio2022}, and others. Pearson correlation captures linear dependency and monotonic non-linear dependencies where variables follow a Gaussian distribution. Spearman correlation handles monotonic non-linear relationships even when variables are not normally distributed. Both Pearson and Spearman correlations are unable to detect non-monotonic relations. Furthermore, distance correlation detects non-linear and non-monotonic dependencies in the data.

Alternatively, correlation metrics based on information theory include the Maximal Information Coefficient (MIC)~\cite{reshef2013}, Kullback-Leibler (KL) divergence~\cite{Kullback51}, and others. These metrics have a foundation in entropy~\cite{Shannon1948} and mutual information between variables. Furthermore, Gavin's work~\cite{Crooks2024} provides several estimations of entropy-based metrics. Entropy-based measures work best when data follows a Gaussian distribution, ensuring that the entropy is directly proportional to the variance. On the contrary, in non-Gaussian settings, variance and entropy have no direct relationship~\cite{Petty2018}. Entropy-based measures can detect even non-functional dependencies via mutual information. MIC is an equitable metric that measures correlation but is sensitive to noise~\cite{Rainio2022}. In contrast, membership inference attacks~\cite{Reza2017, Rahman2018} provide another approach to measure information leakage.

\section{Background}
\label{background}

Maximal Information Coefficient (MIC)~\cite{Lazarsfeld2022} provides an effective metric for detecting correlations. MIC can identify rare and novel relationships in data. MIC is the maximum mutual information over a constellation of grids over data extents (multiple partitions). As a result, MIC is computationally expensive to estimate. Hence, work~\cite{Lazarsfeld2022} proposed approximations and differentially private versions of MIC. Matrix, $\mathbf{A}$ of dim ($k \times \ell$) with $(i, j) \in [k] \times[\ell]$ has count entries, $\mathbf{A}[i][j]$, per cell on the grid. When each row-sum or column-sum of matrix $\mathbf{A}$ is equal, then we have mass-equipartition. Otherwise, we have range-equipartition. Given matrices, $\mathbf{A}, \mathbf{P}$ $\in \mathbb{R}^{k \times \ell}$ with normalized count $\mathbf{P}[i][j]$, where $\mathbf{P} =\frac{1}{n} \cdot \mathbf{A}$.

The discrete mutual information, $I\left(\left.D\right|_G\right)$ is computed using the provided expression as shown in Equation~\ref{eqn:mutmic}, where $p(i, j)$ is the probability score of (row, col) tuple, ($i$, $j$) of the matrix, $\mathbf{P}$, $p(i, )$ is the probability score summed across row, $i$, of matrix, $\mathbf{P}$, and $p(, j)$ is the probability score summed across column, $j$, of matrix, $\mathbf{P}$.

\noindent\begin{minipage}{.6\linewidth}
\begin{equation}
\label{eqn:mutmic}
I\left(\left.D\right|_G\right)=\sum_{i, j} p(i, j) \log _2 \frac{p(i, j)}{p(i, ) p(, j)} .
\end{equation}
\end{minipage}%
\begin{minipage}{.5\linewidth}
\begin{equation}
\label{eqn:normmutmic}
I^{\star}\left(\left.D\right|_G\right):=\frac{I\left(\left.D\right|_G\right)}{\log _2 \min \{k, \ell\}},
\end{equation}
\end{minipage}

\begin{figure}[htbp]
    \centering
    \includegraphics[scale=0.8]{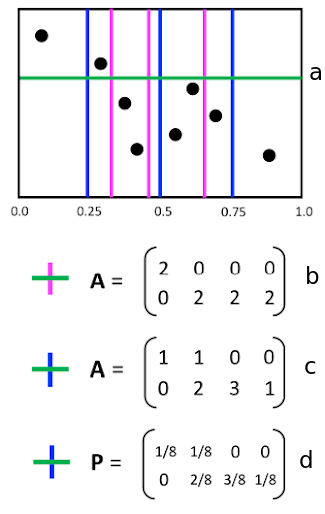}
    \caption{A partitioning of grid~\cite{Lazarsfeld2022} for estimating MIC. Image is organized into sub-charts that are labelled 'a', 'b', 'c' and 'd'}
    \label{fig:mic}

\end{figure}

Discrete mutual information is normalized $I^{\star}\left(\left.D\right|_G\right)$ by using the provided expression as shown in Equation~\ref{eqn:normmutmic}. Let us demonstrate the calculation of normalized mutual information as a procedure for estimating MIC for the grid configuration shown in Figure~\ref{fig:mic}.

It is easier to visualize partitions in 2D as higher dimensions do not make for easy visualization, so in our illustration in Figure~\ref{fig:mic}, We restrict our example to 2D for simplification. Despite the visualization constraint in our chart in Figure~\ref{fig:mic}, the partitioning pattern for MIC extends to higher dimensions. The partitioning in the Figure~\ref{fig:mic} is described as follows:

\begin{itemize}[topsep=7pt, partopsep=0pt, listparindent=0pt, leftmargin=5pt]
\item Chart 'a' as shown in Figure~\ref{fig:mic} has three distinct partitions to split the data across its range. These partitions are colored light pink, light blue, and light green.
\item Chart 'b' shown in Figure~\ref{fig:mic} is a partition colored light pink as seen in Chart 'a' of Figure~\ref{fig:mic}. It displays a count of elements in each grid formed by partitioning.
\item Chart 'c' shown in Figure~\ref{fig:mic} is a partition colored light blue as seen in Chart 'a' of Figure~\ref{fig:mic}. It displays a count of elements in each grid formed by partitioning.
\item Chart 'd' as shown in Figure~\ref{fig:mic} is the normalized count of the values as seen in Chart 'c' of Figure~\ref{fig:mic}. 
\end{itemize}
Let $k = 2, \ell = 4$ for the example depicted in Figure~\ref{fig:mic}, we can estimate $I^{\star}\left(\left.D\right|_G\right) = 0.46688$ and $I^{\star}\left(\left.D\right|_G\right) = 0.46688$.
\begin{definition}\label{mic-definition}
(MIC statistic following Definition 2.1 of~\cite{Lazarsfeld2022})\\ $\operatorname{MIC}(D, B)=\max_{k, \ell: k \ell \leq B(n)}\left(\mathbf{M}_{D}^G\right)_{k, \ell}$ where $B:=B(n)$.
\end{definition}
Following Definition~\ref{mic-definition}, we iterate over permutations of grid configurations that maximize the normalized mutual information, $I^{\star}\left(\left.D\right|_G\right)$.

\section{Method}
\label{extensions}

Information-theoretic~\cite{Shannon1948} approaches to computing mutual information between the original data and randomized data distribution to quantify potential privacy losses. Privacy breaches occur if the randomized output from a privacy-enhancing mechanism can be re-identified to recover the original data (input). MIC is computationally expensive due to the cost of partitioning operations across multiple data dimensions, so we have developed a novel metric for estimating vulnerability to privacy leaks by a combination of degree of anonymity, amplification, and $\rho_1$-to-$\rho_2$ formulation.

This section is structured as follows: a definition of permutation entropy in Subsection~\ref{permutation-entropy}, a description of the degree of anonymity in Subsection~\ref{degree-anonymity}, a relationship between mutual information in Subsection~\ref{relation-amplification}, and a relationship between Amplification and Differential Privacy in Subsection~\ref{amplification-diffprivacy}.

\subsection{ Permutation Entropy}
\label{permutation-entropy}

Permutation entropy, $\mathrm{H}(n)$, is a complexity score capturing the non-redundant measure of information using the intrinsic property of the data. It utilizes the symbolic representation of the data instead of the actual data, $\left(x_1, x_2, \ldots, x_N\right)$, using comparator relations $x_1<x_j$ or $x_1>x_j$, where we estimate the probability of patterns based on ordering relations. Alternatively, if the data is non-numerical, then the data can be encoded using lexical order to impose ordinal relations~\cite{Kozak2020}.

\noindent\begin{minipage}{.5\linewidth}
$$
p(\pi)=\frac{Q(\pi)}{N-n+1} .
$$
\end{minipage}%
\begin{minipage}{.5\linewidth}
$$
\mathrm{H}(n)=-\sum_{i=1}^{n !} p\left(\pi_i\right) \cdot \log \left(\pi_i\right)
$$
\end{minipage}
Where permutation pattern length, $n$, probability of a permutation pattern, $p(\pi)$, $Q(\pi)$ is the frequency of the pattern $\pi$. Permutation entropy can capture complex relations in the data (monotonic, functional, non-functional, linear, non-linear, and others) as part of the ordinal relations. There should be a sufficient length of patterns to capture local dependencies in the data, which may be a crucial factor for privacy vulnerability~\cite{Liu2016}. Unfortunately, it is computationally fast and single-scaled and may require a multi-scale variant~\cite{Ying2022} to capture the most information content.

\subsection{ Degree of Anonymity}
\label{degree-anonymity}

The degree of anonymity~\cite{Koot2023} provides a measure of privacy leaks arising from the probability distribution. Its value is between 0 and 1. Let us define the degree of anonymity, $d$, as $d = \frac {H(P^{*})} {H_M}$, where $H(P^{*})$ is the permutation entropy, $H_M$ is the maximal entropy in the system depicted as $H_M = \log_2(N)$, and $N$ is a number of elements respectively. The characteristics of $d$, are shown as follows: $d = 0$ (attacker succeeds 100\%, predictable) and $d = 1$ (unpredictable, very random).

\subsection{ Relation to Mutual Information}
\label{relation-mutuals}

Mutual information is a metric that quantifies the privacy loss (distribution distance) between original data and randomized data distributions. On the contrary, a mutual information score can be misleading, as privacy breaches can still happen even if the mutual information is small. As a result, amplification is a metric designed to alleviate the deficiencies in quantifying mutual information by providing "worst-case mutual information" with bounds on theoretical privacy breaches.

\subsection{ Relation to Amplification}
\label{relation-amplification}

Amplification, $\gamma$,~\cite{Evfimievski2003} is a metric to quantify privacy leaks without knowledge of the underlying distribution of the original data. This measure limits information leaks by bounding breaches (upper bound by $\gamma$).
\begin{definition}\label{breach-definition}
Following Definition 1 of~\cite{Evfimievski2003}. We can depict a $\rho_1$-to-$\rho_2$ privacy breach with respect to property $Q(x)$ if for some $y \in V_Y$
\noindent\begin{minipage}{.7\linewidth}
$$
\begin{aligned}
\quad \mathbf{P}[Q(X)] \leqslant \rho_1 \text { and } \quad \mathbf{P}[Q(X) \mid Y=y] \geqslant \rho_2 , \\
\end{aligned}
$$
\end{minipage}%
\begin{minipage}{.3\linewidth}
$$
\begin{aligned}
\gamma < \frac{\rho_2}{\rho_1} \cdot \frac{1-\rho_1}{1-\rho_2}
\end{aligned}
$$
\end{minipage}
\end{definition}

Amplification has a bound (0, $ \infty$) and is not equitable and problematic to interpret. Hence, we modify the metric into an equitable metric via the information coefficient of correlation, $r1$ as $r1 = \sqrt{1 - e^{-2 \gamma}} $ with a bound (0, 1). Reinterpret in the light of the changing values of the degree of anonymity in the original data and the noisy response from the randomizer. We simplify Definition~\ref{breach-definition} to consider $Q(x)$ as a randomizer (privacy-preserving transformation) to fit our construction without loss of generality.
\begin{itemize}
\item When $\rho_1$ is smaller than $\rho_2$, then privacy risk decreases
\item When $\rho_1$ is bigger than $\rho_2$, then privacy risk increases
\item When $\rho_1$ is equal to $\rho_2$, then no change in privacy risk
\end{itemize}
We make modifications to $\rho_1$-to-$\rho_2$ to represent the degree of anonymity in the input, $d_1$, and transformed input (noisy response), $d_2$ respectively as shown in Equation~\ref{eqn:deganomgamma}.
\begin{equation}
\label{eqn:deganomgamma}
\gamma < \frac{d_2}{d_1} \cdot \frac{1-d_1}{1-d_2}
\end{equation}

\subsection{ Relationship between Amplification and Differential Privacy}
\label{amplification-diffprivacy}

Differential privacy provides a mechanism for adding noise to data. The resultant transformed data allows public release without compromising the identifiability of individual records.
\begin{definition}\label{diff-definition}
Based on Definition 7 of~\cite{Dwork2017}. (Differential privacy). Given $\epsilon \geq$ 0, a mechanism $\mathcal{A}_{\mathrm{q}}$ is $\epsilon$-differentially private randomizer, $\operatorname{Pr}$ is a probability measure, $\epsilon$ is noise level, ($x, x^{\prime}$) is pair of data points.
$$
\operatorname{Pr}\left[\mathcal{A}_q(x) \in Y\right] \leq e^{\epsilon} \cdot \operatorname{Pr}\left[\mathcal{A}_{q}\left(x^{\prime}\right) \in Y\right] .
$$
\end{definition}
Can we estimate the noise level, $\epsilon$, strictly using an empirical approach and draw a connection to amplification using Definitions~\ref{breach-definition} and~\ref{diff-definition}? Yes, we simplify to obtain the expression as $\epsilon \geq \ln(\gamma)$.

\section{Experiment and Discussions}
\label{discussions}

\begin{figure}[htbp]
    \centering
    \includegraphics[scale=0.6]{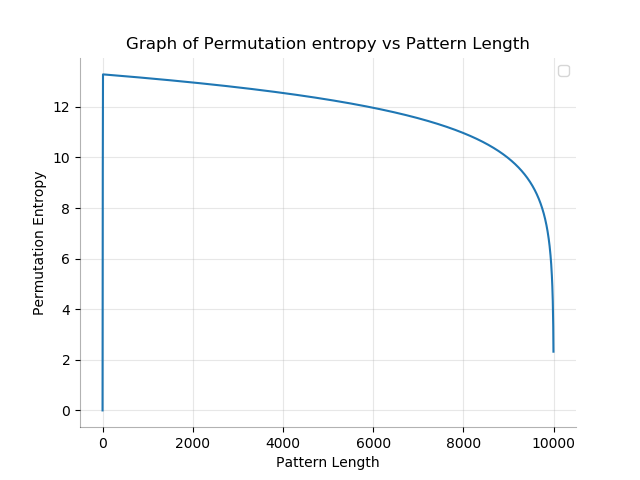}
    \caption{Impact of pattern length on Entropy}
    \label{fig:patternlengthvsentropy}

\end{figure}

We generated a synthetic dataset of 10000 random integers depicting student scores (1, 100) to study the impact of pattern length on permutation entropy. In our experiment, we adopted heuristics for choosing pattern length by taking the minimum pattern length with the maximum entropy values as shown in Figure~\ref{fig:patternlengthvsentropy}. This effect implies an increase in estimated information content as the pattern length increases until it hits a point when increasing pattern length is redundant. Correlation is better estimated using optimal pattern length by doing a hyperparameter search on original data (validation set), using the pattern length to estimate entropy, and comparing outputs of privacy-preserving transformations for susceptibility to privacy leaks. The concavity of permutation entropy as confirmed in Figure~\ref{fig:patternlengthvsentropy}, aligns with the concave property definition of the Shannon entropy~\cite{Shannon1948}.

\begin{figure}[htbp]
    \centering
    \begin{tabular}{|c|c|c|c|c|c|}
    \hline
    \multicolumn{6}{|c|}{Privacy Preserving cases} \\
    \hline
        S/N & Data set & $\rho_1$ & $\rho_2$ & $m$ & $n$\\
    \hline
    1 & Sensor readings & 0.998 & 0.998 & 944 & 11\\
    2 & Customer purchases & 0.999 & 0.999 & 1500 & 10\\
    3 & Employee attrition & 0.999 & 0.999 & 1000 & 9\\
    \hline
    \multicolumn{6}{|c|}{Blatantly Non-Private cases} \\
    \hline
    1 & Sensor readings & 0.998 & 0.0 & 944 & 11\\
    2 & Customer purchases & 0.999 & 0.0 & 1500 & 10\\
    3 & Employee attrition & 0.999 & 0.0 & 1000 & 9\\
    \hline
    \end{tabular}
    \caption{Real-world Dataset Evaluation}
    \label{table:dataanom}

\end{figure}

The degree of anonymity, $\rho_1$, is estimated for the original data as 0.9998 with permutation entropy as 13.29 bits. Also, the randomizer uses an exponential mechanism with privacy-preserving transformed data having the degree of anonymity, $\rho_2$, as 0.9998. Using $\rho_1$-to-$\rho_2$ formulation and pattern length set to 11 for permutation entropy. Consider the values of $\rho_1$, and $\rho_2$ respectively. Since both degrees of anonymity (original and randomized data) are almost equal, we can conclude there is no privacy leak relative to the randomized output. The value of $\rho_1$ is high because we used random numbers as the original data in our demonstration.

We further demonstrate the real-world usefulness by performing anonymization using a differential privacy scheme (exponential mechanism with $\epsilon = 0.337, \delta = 0.1$) on a set of real-world datasets with the size, $m$, and pattern length, $n$ as shown in Figure~\ref{table:dataanom}. Our dataset consists of sensor readings~\footnote{\url{https://www.kaggle.com/datasets/umerrtx/machine-failure-prediction-using-sensor-data}}, customer purchase records~\footnote{\url{https://www.kaggle.com/datasets/rabieelkharoua/predict-customer-purchase-behavior-dataset}}, and employee attrition data~\footnote{\url{https://www.kaggle.com/datasets/mrsimple07/employee-attrition-data-prediction}}. Both the original data and the anonymized data have their dimensionality reduced to 1D using TSNE~\footnote{\url{https://scikit-learn.org/stable/modules/generated/sklearn.manifold.TSNE.html}}. We provide a monotonically increasing sequence of anonymized data for blatantly non-private cases. 

MIC can range from 0 to 1 (low to high correlation). In contrast, the degree of anonymity ranges from 0 to 1 (high to low correlation). Privacy leaks are less likely where there is minimal correlation between the original data and the noisy response from a randomizer. We used the degree of anonymity, $d_1$, $d_2$, as a probability measure that captures the data characteristics without knowing the underlying distribution of the data. Additionally, our choice of pattern length can influence the estimated permutation entropy for a chosen pattern. Alternatively, instead of considering every pattern, we could utilize a restricted set of patterns to estimate the permutation entropy. This choice of pattern creation can incorporate domain knowledge to bias entropy estimates using patterns likely to be identified as potential privacy breaches.

MIC is sensitive to how dataset is partitioned across dimensions, creating regions that can reveal novel correlations and potential privacy vulnerabilities. Unlike our amplification procedure, which requires reducing multivariate data to a one-dimensional representation required for the calculation of the permutation entropy, MIC can handle multidimensional data. Although it is a lossy transform with information loss, it is still reasonable option given that computation uses symbolic representations for this measure. However, both methods (MIC and amplification-based method) are sensitive to grid configuration and pattern length choices. The concept of grid partitioning in MIC is analogous to creating pattern slices for estimating permutation entropy in our amplification scheme. While permutation entropy is computationally efficient, MIC can be more computationally expensive, especially for high-dimensional data, due to the need to maintain partitions of varying sizes. Our entropy-based formulation differs from MIC in that permutation entropy involves one-dimensional slicing, while MIC relies on multidimensional griding.

This work is subject to the inherent limitations of entropy-based measures. For example, permutation entropy relies on arbitrary ordering relations in the symbolic representation of data, which may not be suitable for all applications. Additionally, this structure mandates that we keep the same order in the transformed data to match the order of the original data to enable meaningful comparisons, thereby limiting the flexibility of the method to diverse use cases. Furthermore, the potential correlations with publicly available information may skew the estimation of privacy risk, thereby impacting the interpretability of our $\rho_1$-to-$\rho_2$ formulation.

\section{Conclusions and Future Work}
\label{conclude}

Our amplification formulation ($\rho_1$-to-$\rho_2$) and MIC can identify correlation dependencies between attributes in the data set as a proxy for privacy leak vulnerabilities. Hence, we can utilize the degree of anonymity with the $\rho_1$-to-$\rho_2$ formulation to check susceptibility to privacy leaks among several privacy-preserving systems under evaluation. Furthermore, we can extend this work to detecting privacy leaks in cases (for example, in time series data with autoregressive properties) where conditional permutation entropy~\cite{Gutjahr2021} can better capture intrinsic information than permutation entropy. Finally, we have demonstrated in Figure~\ref{table:dataanom} how to utilize $\rho_1$-to-$\rho_2$ formulation to quantify the privacy risk of the underlying privacy-preserving mechanism.

\bibliographystyle{IEEEtran}
\bibliography{IEEEabrv,sample-sigconf}

\begin{thebibliography}{10}
\providecommand{\url}[1]{#1}
\csname url@samestyle\endcsname
\providecommand{\newblock}{\relax}
\providecommand{\bibinfo}[2]{#2}
\providecommand{\BIBentrySTDinterwordspacing}{\spaceskip=0pt\relax}
\providecommand{\BIBentryALTinterwordstretchfactor}{4}
\providecommand{\BIBentryALTinterwordspacing}{\spaceskip=\fontdimen2\font plus
\BIBentryALTinterwordstretchfactor\fontdimen3\font minus
  \fontdimen4\font\relax}
\providecommand{\BIBforeignlanguage}[2]{{%
\expandafter\ifx\csname l@#1\endcsname\relax
\typeout{** WARNING: IEEEtran.bst: No hyphenation pattern has been}%
\typeout{** loaded for the language `#1'. Using the pattern for}%
\typeout{** the default language instead.}%
\else
\language=\csname l@#1\endcsname
\fi
#2}}
\providecommand{\BIBdecl}{\relax}
\BIBdecl

\bibitem{Liu2016}
C.~Liu, S.~Chakraborty, and P.~Mittal, ``{Dependence Makes You Vulnberable:
  Differential Privacy Under Dependent Tuples},'' in \emph{{Proceedings of the
  Network and Distributed System Security Symposium}}, 2016.

\bibitem{Evfimievski2003}
A.~Evfimievski, J.~Gehrke, and R.~Srikant, ``{Limiting Privacy Breaches in
  Privacy Preserving Data Mining},'' in \emph{{Proceedings of the Symposium on
  Principles of Database Systems}}, 2003, pp. 211--222.

\bibitem{Reshef2011}
D.~N. Reshef, Y.~A. Reshef, H.~K. Finucane, S.~R. Grossman, G.~McVean, P.~J.
  Turnbaugh, E.~S. Lander, M.~Mitzenmacher, and P.~C. Sabeti, ``{Detecting
  novel associations in large data sets},'' \emph{{Science}}, vol. 334, no.
  6062, pp. 1518--1524, 2011.

\bibitem{Lazarsfeld2022}
J.~Lazarsfeld, A.~Johnson, and E.~Adeniran, ``{Differentially Private Maximal
  Information Coefficients},'' in \emph{{Proceedings of the International
  Conference on Machine Learning}}, 2022, pp. 12\,126--12\,163.

\bibitem{Biswas2023}
S.~Biswas, A.~Fole, N.~Khare, and P.~Agrawal, ``{Enhancing correlated big data
  privacy using differential privacy and machine learning},'' \emph{{Journal of
  Big Data}}, vol.~10, no.~30, 2023.

\bibitem{Kifer2014}
D.~Kifer and A.~Machanavajjhala, ``{Pufferfish: A Framework for Mathematical
  Privacy Definitions},'' \emph{{ACM Transactions on Database Systems}},
  vol.~39, no.~1, 2014.

\bibitem{Kozak2020}
J.~Kozak, K.~Kania, and P.~Juszczuk, ``{Permutation Entropy as a Measure of
  Information Gain/Loss in the Different Symbolic Descriptions of Financial
  Data},'' \emph{Entropy}, vol.~22, no.~3, 2020.

\bibitem{Rainio2022}
O.~Rainio, ``{Different Coefficients for Studying Dependence},'' \emph{Sankhya
  B}, vol.~84, no.~2, 2022.

\bibitem{reshef2013}
\BIBentryALTinterwordspacing
D.~Reshef, Y.~Reshef, M.~Mitzenmacher, and P.~Sabeti, ``{Equitability Analysis
  of the Maximal Information Coefficient, with Comparisons},'' 2013. [Online].
  Available: \url{https://arxiv.org/abs/1301.6314}
\BIBentrySTDinterwordspacing

\bibitem{Kullback51}
S.~Kullback and R.~A. Leibler, ``{On Information and Sufficiency},'' \emph{{The
  Annals of Mathematical Statistics.}}, vol.~22, no.~1, pp. 79--86, 1951.

\bibitem{Shannon1948}
C.~Shannon, ``{A Mathematical Theory of Communication},'' \emph{{The Bell
  System Technical Journal}}, vol.~27, pp. 379--423, 1948.

\bibitem{Crooks2024}
G.~E. Crooks, ``{On Measures of Entropy and Information},''
  \url{https://threeplusone.com/pubs/on_information.pdf}, 2024, {Date accessed:
  April 10, 2024}.

\bibitem{Petty2018}
G.~W. Petty, ``{On Some Shortcomings of Shannon Entropy as a Measure of
  Information Content in Indirect Measurements of Continuous Variables},''
  \emph{{Journal of Atmospheric and Oceanic Technology}}, vol.~35, no.~5, pp.
  1011--1021, 2018.

\bibitem{Reza2017}
R.~Shokri, M.~Stronati, C.~Song, and V.~Shmatikov, ``{Membership Inference
  Attacks Against Machine Learning Models},'' in \emph{{Symposium on Security
  and Privacy}}, 2017, pp. 3--18.

\bibitem{Rahman2018}
M.~Rahman, T.~Rahman, R.~Lagani{\`e}re, and N.~Mohammed, ``{Membership
  Inference Attack against Differentially Private Deep Learning Model},''
  \emph{{Transactions on Data Privacy}}, vol.~11, pp. 61--79, 2018.

\bibitem{Ying2022}
W.~Ying, J.~Tong, Z.~Dong, H.~Pan, Q.~Liu, and J.~Zheng, ``{Composite
  Multivariate Multi-Scale Permutation Entropy and Laplacian Score Based Fault
  Diagnosis of Rolling Bearing},'' \emph{{Entropy}}, vol.~24, no.~2, 2022.

\bibitem{Koot2023}
M.~R. Koot, ``{Measuring and Predicting Anonymity},''
  \url{https://pure.uva.nl/ws/files/1834030/107610_thesis.pdf}, 2012, {Date
  accessed: July 28, 2023}.

\bibitem{Dwork2017}
C.~Dwork, A.~D. Smith, T.~Steinke, and J.~Ullman, ``{Exposed! A Survey of
  Attacks on Private Data},'' \emph{{Annual Review of Statistics and Its
  Application}}, vol.~4, pp. 61--84, 2017.

\bibitem{Gutjahr2021}
T.~Gutjahr and K.~Keller, ``{Conditional Permutation Entropy as a measure for
  the Complexity of Dynamical Systems},'' in \emph{{Proceedings of the Entropy:
  The Scientific Tool of the 21st Century}}, 2021, pp. 12\,126--12\,163.

\end{thebibliography}

\newpage
\section{Appendix}

\subsection{Threat Model}

The attacker may have access to the original data before randomization and may want to know how much information can be gleaned by observing the outputs of the privacy-preserving mechanism. The attacker has observed both the original and randomized data. Reconstruction attacks are possible by finding the relationship between the original data and the output from the privacy-preserving transformed output, with the order preserved (unshuffled) in both instances (original and private data). We want to account for the potential of privacy leaks due only to the effect of the privacy-preserving algorithm without shuffling (order-preserving).

\subsection{Proof of susceptibility to privacy leaks}

Using ideas from Theorem 1 of ~\cite{Shannon1948} that proved a mathematical definition of Shannon Entropy. We have adapted its construction in the formulation of permutation entropy as shown in Subsection~\ref{permutation-entropy}. The generic Shannon entropy representation is a specialization of our definition of permutation entropy.

Each probability event in the entropy formulation is a set of patterns. This entropy estimation scheme has a desirable characteristic that is compatible with the symbolic data representation in our formulation for modeling data-agnostic representation. Another benefit of using a set of patterns for probability events is the flexibility permitting us to arbitrarily modify weights of known patterns that have increased privacy leak susceptibility.

Following the property of maximal entropy is $log_b n$, where $p_1, \ldots, p_n$ are $n$ probability events, and is the entropy, $\mathrm{H}$, then the following condition holds
$$
\mathrm{H}\left(p_1, \ldots, p_n\right) \leq \log_b n
$$
Hence, we have used the maximal entropy idea is used to normalize the entropy, $\mathrm{H} \in [0, 1]$ which known as the degree of anonymity, $d$, depicted in Subsection~\ref{degree-anonymity} and Equation~\ref{eqn:deganomityproof}.
\begin{equation}
\label{eqn:deganomityproof}
d = \frac {H(P^{*})} {H_M}
\end{equation}
where $H(P^{*})$ is the permutation entropy, $H_M$ is the maximal entropy in the system depicted as $H_M = \log_2(N)$, and $N$ is a number of elements respectively.

The symmetry of the entropy measure is shown in Equation~\ref{eqn:symetry}.
\begin{equation}
\label{eqn:symetry}
\mathrm{H}(X, Y)=\mathrm{H}(X \mid Y)+\mathrm{H}(Y)=\mathrm{H}(Y \mid X)+\mathrm{H}(X)
\end{equation}

Simplifying Equation~\ref{eqn:symetry}, we have the both expressions shown in Equation~\ref{eqn:simpsymetry1}
$$\mathrm{H}(X \mid Y) = \mathrm{H}(Y) - \mathrm{H}(X, Y) $$ 
Or
\begin{equation}
\label{eqn:simpsymetry1}
    \begin{split}
        \mathrm{H}(Y \mid X) =\mathrm{H}(X) - \mathrm{H}(X, Y)
    \end{split}
\end{equation}
If $X, Y$ are independent then the following condition holds in Equation~\ref{eqn:simpsymetry2}
\begin{equation}
\label{eqn:simpsymetry2}
    \begin{split}
        \mathrm{H}(Y \mid X) =\mathrm{H}(X)
    \end{split}
\end{equation}
The implication of Equation~\ref{eqn:simpsymetry2} is that if consider $Y$ related to $X$ by a function, f say $Y= f(X)$ where $f$ is the privacy-preserving transform when both variables ($X, Y$) are independent, we cannot derive information about $X$ from $Y$. Hence, there is less susceptibility to a privacy leak. Given two events ($X, Y$) following properties of Shannon entropy, the condition holds
$\mathrm{H}(X, Y) \leq \mathrm{H}(X)+\mathrm{H}(Y)$ with equality of the relation when ($X, Y$) are independent.

Let us use the dependence property to prove that our adaptation of $\rho_1$-to-$\rho_2$ formulation captures privacy risk. The proof covers four main cases.

\vspace{15pt}
\textbf{Case 1}: Following Equation~\ref{eqn:simpsymetry1}, when $\mathrm{H}(X, Y) \approx 0$ may arise due to Negative Exponential Law states as the frequency of independent event increases, the joint probability tends to zero. It can happen when both events (original data $X$, private derived data $Y$) are independent. We obtain the condition where there is no change in privacy risk. The interpretation is that the privacy-preserving transformation does not impact the privacy risk in the private data relative to the privacy susceptibility in the original data.

\vspace{10pt}
\textbf{Case 2}: Following Equation~\ref{eqn:simpsymetry1}, when $\mathrm{H}(X, Y) > 0$ or ($\mathrm{H}(X) > \mathrm{H}(X, Y)$) which can happen when both events (original data $X$, private derived data $Y$) are "almost" independent. We obtain the condition where there is a decrease in privacy risk. The interpretation is that the privacy-preserving transformation decreases the privacy risk in the private data relative to the privacy susceptibility in the original data.

\vspace{10pt}
\textbf{Case 3}: Following Equation~\ref{eqn:simpsymetry1}, when $\mathrm{H}(X, Y) < 0$ can happen when both events (original data $X$, private derived data $Y$) are dependent. We obtain the condition where there is an increase in privacy risk. The interpretation is that the privacy-preserving transformation increases the privacy risk in the private data relative to the privacy susceptibility in the original data.

\vspace{10pt}
\textbf{Case 4}: Following Equation~\ref{eqn:simpsymetry1}, when $\mathrm{H}(X, Y) = \mathrm{H}(Y)$, then $\mathrm{H}(X \mid Y) =0$ (not applicable in our formulation) and when $\mathrm{H}(X, Y) = \mathrm{H}(X)$, then $\mathrm{H}(Y \mid X) =0$. The interpretation of these relations is that there is no relationship between both events (original data $X$, private derived data $Y$). 

\vspace{10pt}
\textbf{Deduction}: Thus, we have demonstrated that our adaptation of $\rho_1$-to-$\rho_2$ formulation utilizing entropy, mutual information, and degree of anonymity is a reasonable measure of privacy risk where $\rho_1$ is the degree of anonymity on the original data, $X$, and $\rho_2$ is the degree of anonymity of the privately derived data $Y$. $Y = f(X)$ where $f$ is the privacy-preserving mechanism. The degree of anonymity is a normalized entropy, $\mathrm{H}$ utilizing the maximal entropy as noted in Subsection~\ref{degree-anonymity}.

\end{document}